\documentclass[11pt,amsmath,preprintnumbers,amssymb,aps,floatfix,superscriptaddress]{revtex4-1}

\pdfoutput=1

\usepackage{epsfig}
\usepackage{graphicx}
\usepackage{graphicx}
\usepackage{dcolumn}
\usepackage{bm}
\usepackage{hyperref}

\begin{document}
\title{Coherent properties of single rare-earth spin qubits.}
\author{P.~Siyushev}
\affiliation{3. Physikalisches Institut, Universit\"at Stuttgart and Stuttgart Research Center of Photonic Engineering (SCoPE), Pfaffenwaldring 57, Stuttgart, D-70569, Germany}
\author{K.~Xia}
\affiliation{3. Physikalisches Institut, Universit\"at Stuttgart and Stuttgart Research Center of Photonic Engineering (SCoPE), Pfaffenwaldring 57, Stuttgart, D-70569, Germany}
\author{R.~Reuter}
\affiliation{3. Physikalisches Institut, Universit\"at Stuttgart and Stuttgart Research Center of Photonic Engineering (SCoPE), Pfaffenwaldring 57, Stuttgart, D-70569, Germany}
\author{M.~Jamali}
\affiliation{3. Physikalisches Institut, Universit\"at Stuttgart and Stuttgart Research Center of Photonic Engineering (SCoPE), Pfaffenwaldring 57, Stuttgart, D-70569, Germany}
\author{N.~Zhao}
\affiliation{Beijing Computational Science Research Center, Beijing 100084, China}
\author{N.~Yang}
\affiliation{Institute of Applied Physics and Computational Mathematics, P.O. Box 8009(28),100088 Beijing, China}
\author{C.~Duan}
\affiliation{Hefei National Laboratory for Physics Sciences at Microscale and Department of Physics, University of Science and Technology of China, Hefei, 230026, China}
\author{N.~Kukharchyk}
\affiliation{Ruhr-Universit\"at Bochum, Universit\"atsstra\ss e 150 Geb\"aude NB, D-44780 Bochum, Germany}
\author{A.D.~Wieck}
\affiliation{Ruhr-Universit\"at Bochum, Universit\"atsstra\ss e 150 Geb\"aude NB, D-44780 Bochum, Germany}
\author{R.~Kolesov}
\affiliation{3. Physikalisches Institut, Universit\"at Stuttgart and Stuttgart Research Center of Photonic Engineering (SCoPE), Pfaffenwaldring 57, Stuttgart, D-70569, Germany}
\author{J.~Wrachtrup}
\affiliation{3. Physikalisches Institut, Universit\"at Stuttgart and Stuttgart Research Center of Photonic Engineering (SCoPE), Pfaffenwaldring 57, Stuttgart, D-70569, Germany}

\begin{abstract}
Rare-earth-doped crystals are excellent hardware for quantum storage of optical information \cite{storage1,storage2,high_BW_storage}. Additional functionality of these materials is added by their waveguiding properties \cite{ceramic_waveguides_YAG,ceramic_waveguides_ruby} allowing for on-chip photonic networks. However, detection and coherent properties of rare-earth single-spin qubits have not been demonstrated so far. Here, we present experimental results on high-fidelity optical initialization, effcient coherent manipulation, and optical readout of a single electron spin of $Ce^{3+}$ ion in a YAG crystal. Under dynamic decoupling \cite{DD}, spin coherence lifetime reaches $T_2=2\;ms$ and is almost limited by the measured spin-lattice relaxation time $T_1=3.8\;ms$. Strong hyperfine coupling to aluminium nuclear spins suggests that cerium electron spins can be exploited as an interface between photons and long-lived nuclear spin memory. Combined with high brightness of $Ce^{3+}$ emission and a possibility of creating photonic circuits out of the host material, this makes cerium spins an interesting option for integrated quantum photonics.
\end{abstract}

\maketitle

Outstandingly long spin coherence times of rare-earth-based optical crystals result in long-lasting memories for storing quantum states of single photons \cite{photon_memory} as well as the states of entangled photon pairs \cite{photon_entanglement1,photon_entanglement2}. In these quantum memories the quantum state of a photon is stored in an electron or nuclear spin wave created in an ensemble of spins. This type of memory is an essential ingredient of quantum repeaters \cite{quantum_repeater} and quantum computing protocols based on linear optics \cite{linear_optics}. However, ensemble-based rare-earth quantum computer designs lack scalability \cite{EuYSO_conditional_phase_shift} in a sense that the number of interacting qubits is stricktly limited by the number of spin states. This problem can be resolved by addressing single rare-earth spins individually. In the last two years, detection of several rare-earth species in a crystal host at a single ion level was demonstrated \cite{PrYAG,ErSi,CeYAG} though no long-lasting spin coherence was demonstrated so far. In this Letter, we present for the first time optical initialization, coherent manipulation, and optical readout of a single cerium ion in a crystalline host.

\begin{figure}[t]
\center{
\includegraphics[width=13cm]{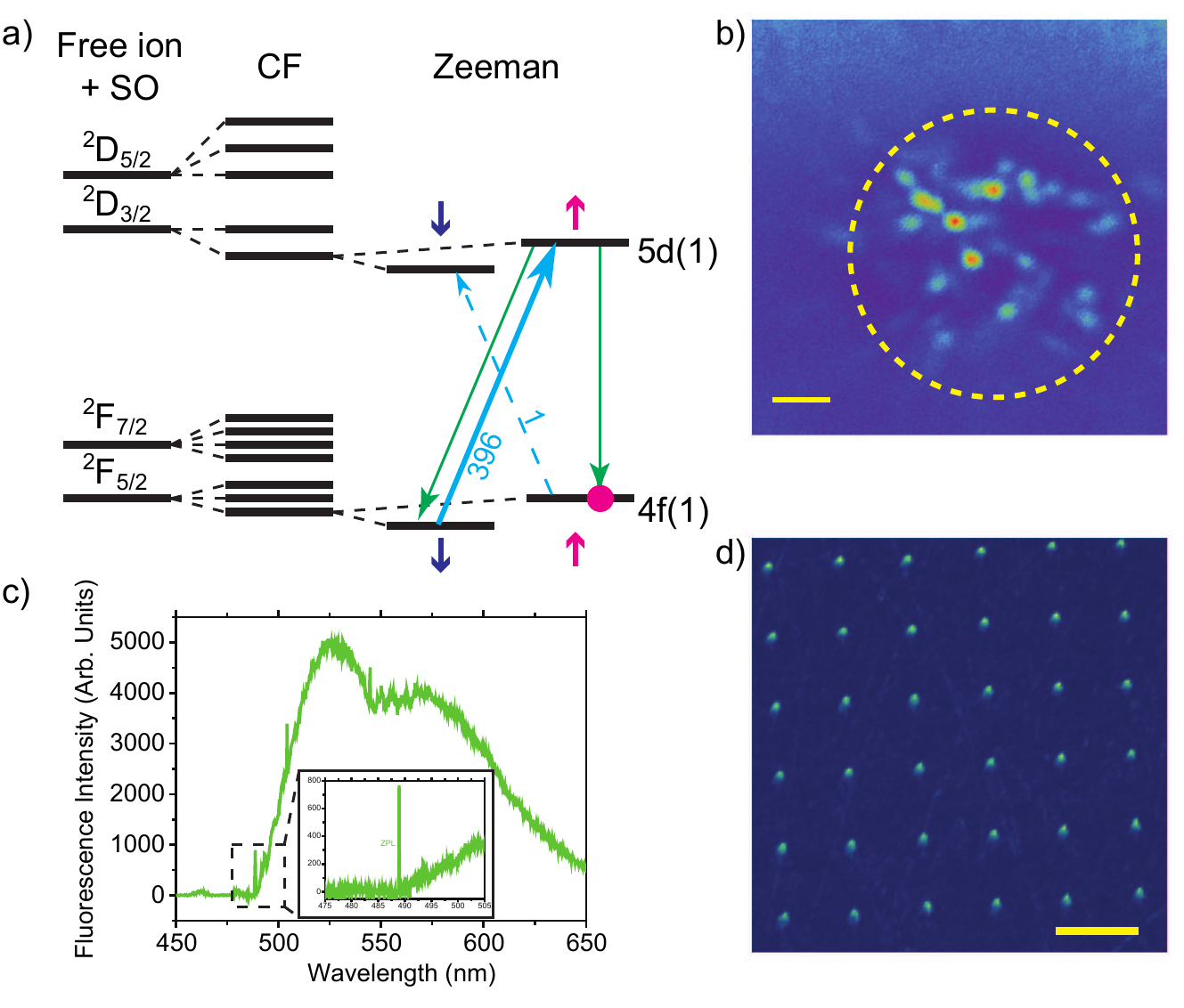}
\caption{\label{fig:levels} {\bf General properties of a single $Ce^{3+}$ ion in YAG.} a) Electronic level structure of $Ce^{3+}$ ion in YAG crystal. Relative strengths of optical transitions between the lowest 4f and the lowest 5d spin doublets (396:1) are shown for circularly polarized excitation light propagating along the y-axis of the local $Ce^{3+}$ symmetry frame. External magnetic field along the same direction splits the spin doublets. b) Confocal scan of individual cerium centres under the SIL. The scale bar is $2\;\mu m$. The dashed circle indicates the boundary of the SIL. c) Typical spectrum of a single cerium ion. The inset shows a zoom into the ZPL of $Ce^{3+}$. The positions of ZPLs for distinct cerium centres are slightly different depending on the local environment of each centre. d)  Focused beam of cerium ions accelerated to $300\;keV$ was used to create a grid of fluorescent spots in YAG single crystal. The scale bar is $10\;\mu m$. Measurements of fluorescence lifetime, spectra, and excited state Larmor precession proved that the fluorescence originates from substitutional $Ce^{3+}$ ions in YAG.}
}
\end{figure}

Experimentally, individual cerium ions were identified in an ultra-pure YAG crystal by confocal microscopy \cite{CeYAG} at $T\approx 3.5\;K$ (see Methods section and \href{http://dx.doi.org/10.1038/ncomms4895}{Supplementary Figure S1}). In order to improve the fluorescence collection efficiency of the microscope, a solid immersion lens (SIL) was fabricated on the surface of the sample by focused ion beam milling. Cerium emission associated with the $5d\rightarrow 4f$ (see Figure \ref{fig:levels}a) transition was excited by a frequency-doubled femtosecond Ti:Sapphire laser at $460\;nm$ and detected in $475-630\;nm$ spectral window. The image of individual cerium ions under the SIL is shown in Figure \ref{fig:levels}b. The detected emission of an individual cerium ion amounts to $60-70\times 10^3\;photons/s$. The emission spectrum of the ions is indicated in Figure \ref{fig:levels}c showing broadband phonon-assisted emission and a sharp zero-phonon line at $489\;nm$ characteristic for cerium ions \cite{CeYAG_ZPL}. Cerium fluorescent centres can be created in YAG crystal by ion implantation (see Figure \ref{fig:levels}d). Comparing the brightness of a single cerium ion and that of the implanted regions and assuming the implantation dose to be known (though the dose is believed to be significantly overestimated), we obtained the lower boundary of the implantation yield of $8\%$.
The electronic level structure of cerium substitutional impurities in ceramic crystals is shown in Figure \ref{fig:levels}a. The strongest transition between the ground 4f(1) and the optically excited 5d(1) spin doublets under $\sigma^+$ circularly polarized excitation is the a spin flip transition $\left|4f(1)\downarrow\right\rangle\rightarrow\left|4d(1)\uparrow\right>$. The other 3 transitions are almost 3 orders of magnitude weaker \cite{CeYAG}. However, the decay back to the ground state is equally probable. This results in efficient pumping of the ion into spin-up ``dark'' state. In this way, cerium spin state can be initialized.
\begin{figure}[b]
\center{
\includegraphics[width=13cm]{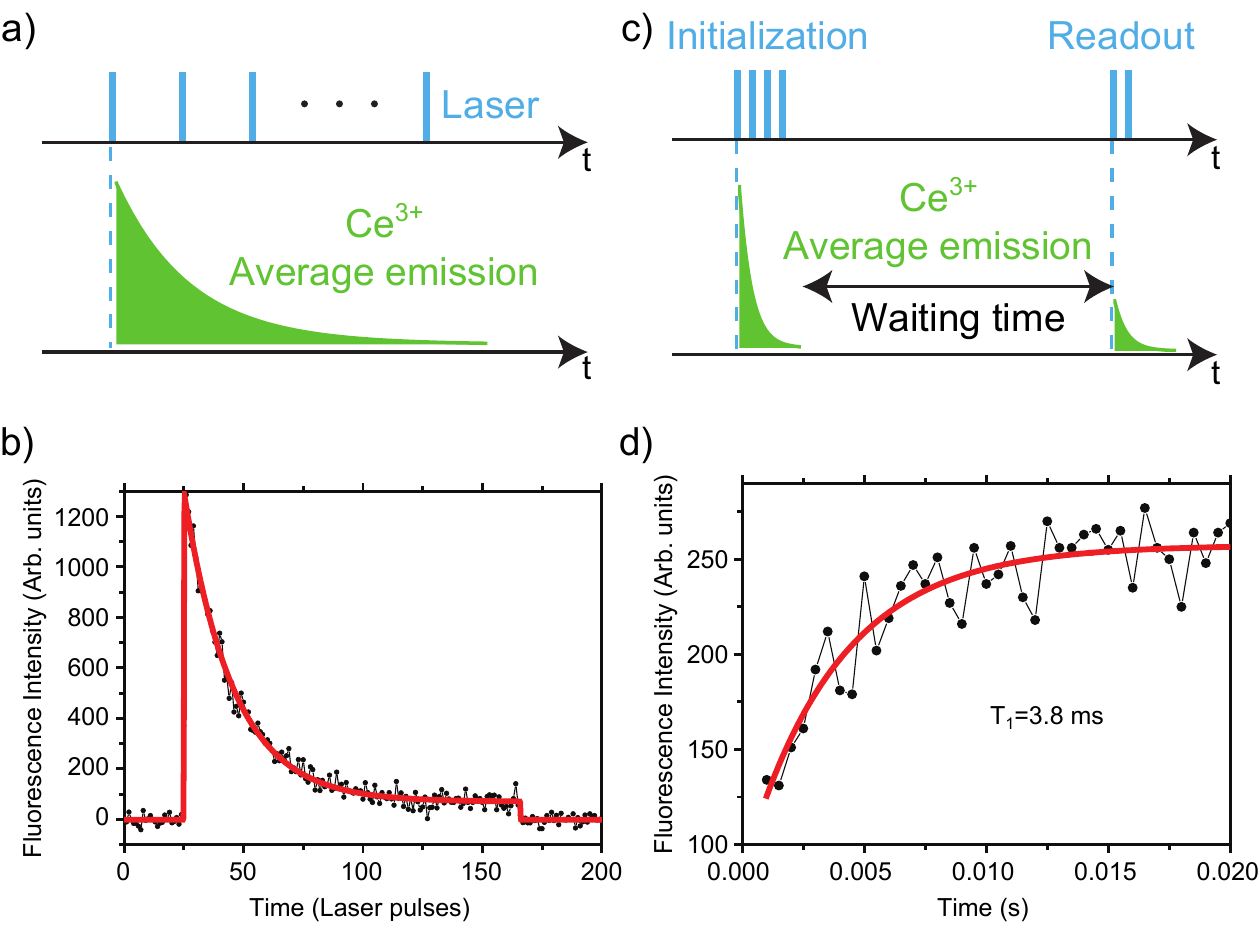}
\caption{\label{fig:initialization} {\bf Initialization and thermalization of $Ce^{3+}$ spin.} a) Schematic diagram of the laser pulse sequence used to study the initialization dynamics. The time interval between the trains was chosen to be long enough for the spin population to thermalize. The photons accumulated after each pulse in the train are summed up and represent one point on the initialization fluorescence curve. b) Experimental dependence of the fluorescence on time during initialization process. c) Laser pulse sequence used to measure spin $T_1$ time. The initializing pulse train is long (around 50 pulses) to ensure spin polarization. The readout pulse train contains only 5 pulses. The time interval between initializing the spin and reading it out is varied. d) Spin relaxation curve recorded by detecting the fluorescence during the readout pulse train.}
}
\end{figure}
To reveal pumping dynamics of cerium spin, magnetic field of $490\;G$ parallel to the excitation beam was applied to the sample, the excitation laser was chopped such that the cerium centre is excited with a finite train of pulses and the time evolution of the excited fluorescence was investigated (see Figure \ref{fig:initialization}a). As shown in Figure \ref{fig:initialization}b, right after the circularly polarized excitation is switched on, the emission intensity is maximal determined by thermalized populations of the ground-state spin sublevels. It rapidly drops due to optical pumping into the ``dark'' spin state. 
The ratio of the fluorescence intensities in the beginning of the fluorescence decay and after the spin is polarized 
indicates $>97\%$ fidelity of spin initialization. The fidelity of spin initialization not reaching the theoretically predicted ratio $396:1$ is thought to be due to slight ellipticity of the polarization caused by non-perfect $\lambda/4$ waveplate and polarization distortion due to SIL. Under linearly polarized excitation, cerium emission does not depend on time indicating the absence of optical pumping. The dependence of the emission on the ellipticity of the excitation is given in \href{http://dx.doi.org/10.1038/ncomms4895}{Supplementary Figure S2}.

The lifetime of the initialized spin state $T_1$ was measured by introducing the second (readout) train of laser pulses having the same circular polarization and varying the time interval between initialization and readout (see Figure \ref{fig:initialization}c). The fluorescence recovery indicates thermalization of the spin states. The measurement resulted in the spin relaxation time of $T_1=3.8\;ms$ (see Figure \ref{fig:initialization}d). 

\begin{figure}[b]
\center{
\includegraphics[width=13cm]{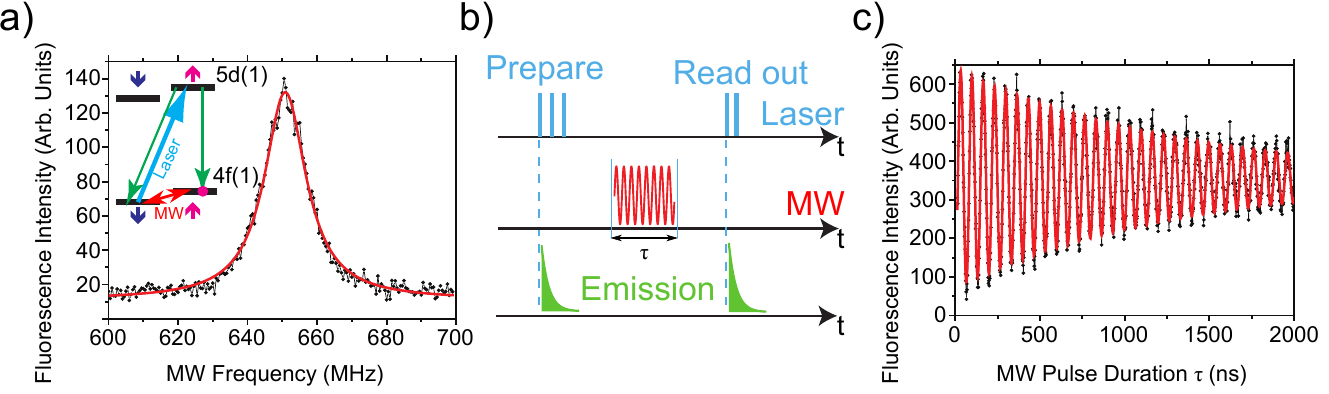}
\caption{\label{fig:ODMR_Rabi} {\bf Coherent manipulation of $Ce^{3+}$ spin.} a) ODMR signal of a single $Ce^{3+}$ ion. Inset shows level structure of the ground-state and the optically excited spin doublets. b) Schematic diagram of Rabi measurements. c) MW Rabi oscillations of a single $Ce^{3+}$ spin.}
}
\end{figure}

The initialized spin can be manipulated by microwave radiation resonant with the spin transition of the ground $4f(1)$ state. For that purpose, a microwave structure was created on the surface of the sample right next to the location of the SIL. This allowed us to record optically detected magnetic resonance (ODMR) spectra of individual cerium ions. Once the frequency of microwaves matches that of the spin transition, the population from the $\left|4f(1)\uparrow\right>$ state is pumped back into $\left|4f(1)\downarrow\right>$ and the overall fluorescence yield increases. Typical ODMR spectrum of a single cerium ion is shown in Figure \ref{fig:ODMR_Rabi}a. Linewidths of ODMR spectra are in the range of $10-15\;MHz$ varying slightly for different cerium centres. Since $Ce^{3+}$ can occupy 6 magnetically inequivalent sites in YAG, in general, 6 distinct positions of ODMR peak are possible.A combination of measured ODMR frequency and of known g-tensor of ground-state $Ce^{3+}$ spin \cite{CeYAG_EPR} allowed us to determine the orientation of individual cerium sites.  In our (110) oriented crystal, the ODMR resonances were at $650\;MHz$, $1310\;MHz$ and four nearly degenerate ones around $1550\;MHz$. For further experiments we were choosing the centres having their local y-axis oriented along the magnetic field and the excitation beam ($640\;MHz$) since for these centres the theoretical predictions for spin initialization fidelity is the best.
The first step towards coherent manipulation of cerium electron spin is detection of Rabi oscillations under strong microwave driving. Experimentally, circularly polarized laser excitation was chopped such that only a few tens of laser pulses are used to initialize the spin in $\left|4f(1)\uparrow\right\rangle$. After this laser pulse train, a microwave pulse at resonant frequency of variable duration was applied (see Figure \ref{fig:ODMR_Rabi}b). Finally, a short laser pulse train consisting of 2-5 pulses having the same circular polarization was applied to read out the population of spin state $\left|4f(1)\downarrow\right\rangle$. The result of the measurement is shown in Figure \ref{fig:ODMR_Rabi}c. The frequency of Rabi oscillations has square-root dependence on the microwave power. The decay time of oscillations is longer than that determined by the inhomogeneous linewidth indicating partial locking of the spin by microwaves.

\begin{figure}[t]
\center{
\includegraphics[width=13cm]{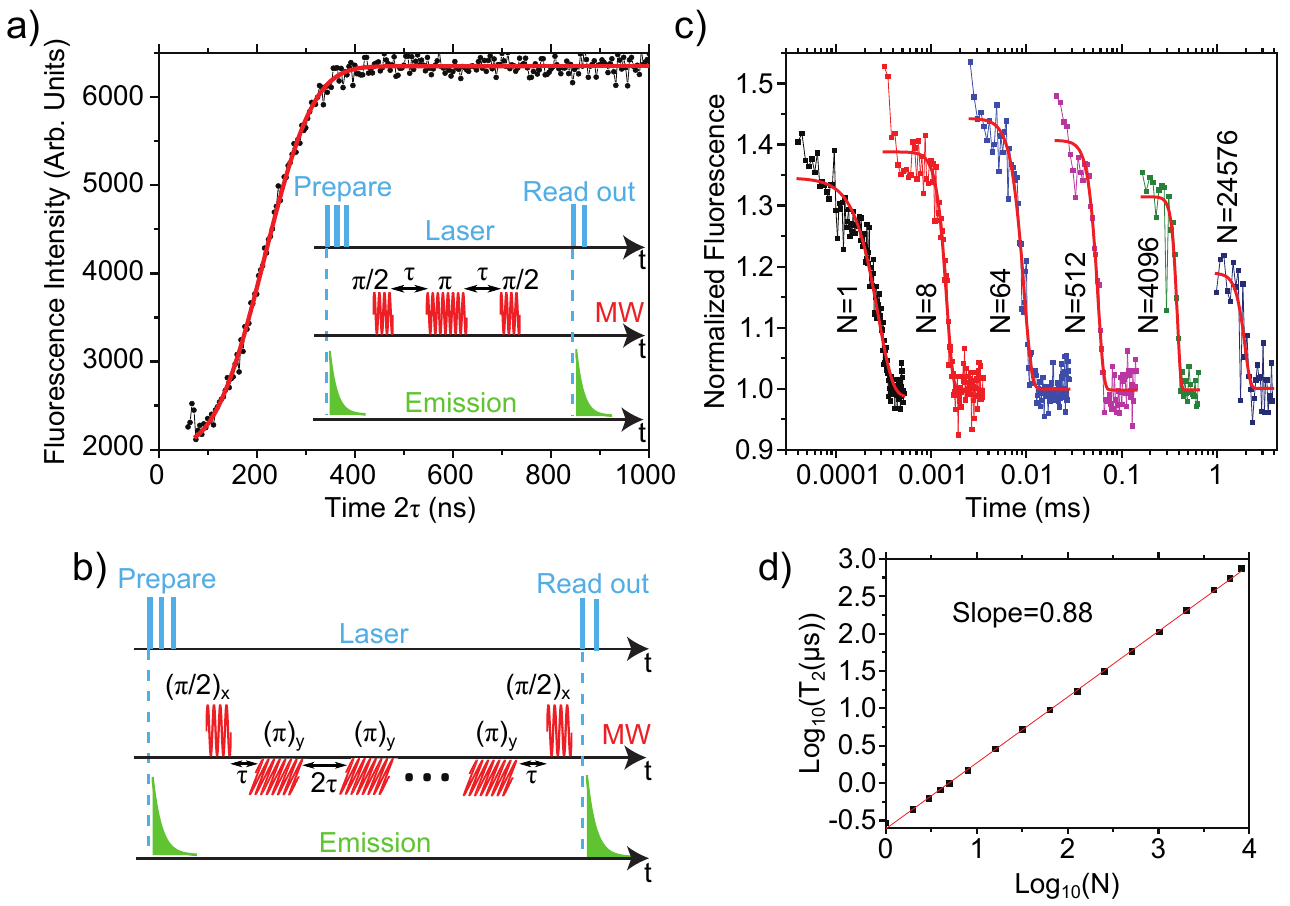}
\caption{\label{fig:echo_CPMG} {\bf Dynamic decoupling of $Ce^{3+}$ spin from the spin bath.} a) Spin-echo decay shows the decoherence time $T_2=240\;ns$. The decay is non-exponential. b) Schematic diagram of measuring the decoherence time with CPMG decoupling sequence. After applying the first $\pi/2$ microwave pulse (X $\pi/2$-pulse) and waiting for time $\tau$, one applies a sequence of $\pi$-microwave refocusing pulses separated by $2\tau$ from each other and phase-shifted from the X $\pi/2$-pulse by $90^{\circ}$ (Y $\pi$-pulses). Rotation in the Y- instead of the X-plane on the Bloch sphere makes the sequence robust against slight deviations of the refocusing pulses from $\pi$ area. Finally, the degree of coherence can be tested by applying a second X $\pi/2$-pulse and reading out the population in the ``bright'' spin state with the readout laser pulse train. c) Results of CPMG measurements for several sequences. The number of pulses for each sequence is given next to the corresponding decay curve. d) The dependence of the coherence lifetime on the length of CPMG sequence. The slope of 0.88 is closer to the linear dependence characteristic to nuclear spin bath rather than to $N^{2/3}$ characteristic to the electron spin bath.}
}
\end{figure}

Dephasing of cerium spin was determined by measuring the decay of Hahn echo signal (a sequence of $\pi/2-\tau-\pi-\tau-\pi/2$ microwave pulses, see inset in Figure \ref{fig:echo_CPMG}a). The result yields a value of $T_2=2\tau=240\;ns$. Fast decoherence in YAG can be explained by the presence of the dense bath of $^{27}Al$ having spin of $I=5/2$ accompanied by large magnetic moment $\mu_{Al}=3.64\mu_N$ with $\mu_N$ being nuclear magneton. Fluctuating nuclear spins produce magnetic noise at the location of $Ce^{3+}$ reducing the coherence time. Theoretical analisys of the nuclear spin bath showed that the contribution of octahedrally coordinated $^{27}Al$ nuclei to decoherence is the largest (see \href{http://dx.doi.org/10.1038/ncomms4895}{Supplementary Methods}). 

Dynamic decoupling exploiting sequences of $\pi$ microwave pulses allows for enormous increase in decoherence time. Decoherence is caused by the averaging over a phase factor $C\left(t\right)=\left\langle e^{i\varphi\left(t\right)}\right\rangle$ the $Ce^{3+}$ spin acquires during free evolution. $\pi$ pulses applied with frequency $1/\tau$ effectively eliminate noise spectral components $\nu_{noise}<1/\tau$ and thus lead to lenghtening of coherence lifetime.
We used Carr-Purcell-Meiboom-Gill (CPMG) microwave decoupling sequence \cite{CPMG1,CPMG2} (see Figure \ref{fig:echo_CPMG}b).   The maximum number of refocusing pulses applied was 24576. This allowed us to extend decoherence time to $1.97\;ms$. The measured CPMG-echo signals shown in Figure \ref{fig:echo_CPMG}c agree well with the ones simulated by correlated cluster expansion method \cite{CCE} (see \href{http://dx.doi.org/10.1038/ncomms4895}{Supplementary Figure S3}). The dependence of the decoherence time on the number of CPMG pulses allows one to discriminate the influence of nuclear and electronic spin baths. The latter is composed of the spins of residual transition metal impurities such as $Cr^{3+}$ and $Fe^{3+}$. Electron and nuclear spin baths have significantly different noise spectral densities. The noise of the electron spin bath is best decribed by a lorenzian shaped spectrum characterized by a long spectral tail. This leads to a $N^{2/3}$ depencence of the coherence time \cite{N23}. On the other hand, nuclear spin noise is characterized by a much harder cut-off leading to a linear dependence of $T_2$ on $N$. The experimentally measured dependence shown on Figure \ref{fig:echo_CPMG}d is $T_2\propto N^{0.88}$ indicating that both nuclear and electron spin baths lead to decoherence with former contribution prevailing. Stretched exponential fits ($\exp\left(-\left(t/T_2\right)^k\right)$) to CPMG-echo signals result in exponentials $k$ between 6 and 12 confirming the existence of hard cutoff of the nuclear spin noise spectrum as described in \href{http://dx.doi.org/10.1038/ncomms4895}{Supplementary Methods}. These results suggest that changing the host material to a low nuclear spin one, e.g. yttrium orthosilicate (YSO) in which rare-earth ions show the longest coherence and population lifetimes \cite{CeYSO,storage2,EuYSO_holeburning}, should greately increase coherence lifetime. Changing the host material has almost no effect on the atomic-like polarization selection rules of $Ce^{3+}$ crucial for initializing its spin state. Our preliminary measurements show extended coherence lifetime in the excited $5d$ state in YSO compared to the YAG host (see \href{http://dx.doi.org/10.1038/ncomms4895}{Supplementary Figure S4}).

In conclusion, we have demonstrated optically addressable rare-earth spin qubit in a ceramic crystal. Technologically advantageous feature of these materials is a possibility of creating them in a form suitable for integrated photonic circuit fabrication, e.g.  in the form of thin films on low refractive index substrates (quartz). Subsequent FIB milling would allow for creating on-chip photonic circuits with low attenuation ($1.5\;dB/cm$ \cite{ceramic_waveguides_YAG}) for single photons which would include waveguides, cavities, etc. This would result in technologically simple realization of a lab-on-chip approach for cerium spin qubits and, simultaneously, in obtaining an access to longer-living isolated nuclear spins magnetically coupled the electron spin of $Ce^{3+}$.

\section*{Methods}

\subsection*{Experimental Setup}

Optical studies of single $Ce^{3+}$ ions were performed in a home-built confocal microscope in which the crystal was mounted on a cold finger of a helium flow cryostat. An optical access was arranged through a window. The microscope objective lens of NA 0.95 was mounted inside the cryostat on a piezo nano-positioner. A toroidal permanent magnet was fixed onto the objective lens to provide magnetic field parallel to the propagation direction of the excitation laser beam. The fluorescence of cerium ions was collected by the same objective lens and detected with a single photon counting avalanche photodiode. In order to measure the emission spectra, the fluorescence was deflected by a mirror mounted on a flip mount onto a grating spectrometer equipped with a cooled CCD camera. For measurements involving microwaves, a high power MW amplifier ($50\;dB$ amplification, $30\;W$ maximum output power) was used. The frequency of microwaves was swept by a software-controlled MW synthesizer.

\subsection*{Sample Preparation}

In order to improve fluorescence collection efficiency, a solid immersion microlens was fabricated directly on the surface of the sample by focused ion beam milling. The lens had a shape of half-a-sphere with the radius of $5\;\mu m$. In order to perform spin manipulations with microwaves, a copper microwave structure was created next to the location of the SIL by lithographic means. Even though the microwave structure was not impedance-matched to the outer MW network, the measured overall MW loss amounted to only $6\; dB$.

\end{document}